\documentclass[pra,a4paper,nofootinbib,twocolumn,showpacs,preprintnumbers,amsmath,amssymb,floatfix,amstex,superscriptaddress]{revtex4}
\newcommand{\ket}[1]{|#1\rangle}                        %
\newcommand{\bra}[1]{\langle #1}                        %
\usepackage{graphicx,graphics,wrapfig,rotating}         
\usepackage{dcolumn}                                    
\usepackage{bbm, bm,fancybox}                           
\usepackage{times,euscript,eufrak,oldgerm}              %
\usepackage[german,english]{babel}                      %
\usepackage{psfrag}
\usepackage{color}

\begin{document}
\title{Scalable experimental estimation of multipartite entanglement}
\author{Leandro Aolita}

\affiliation{%
Instituto de F\'\i sica, Universidade Federal do Rio de Janeiro, Caixa Postal
68528, 21941-972 Rio de Janeiro, RJ, Brasil\\
}
\affiliation{%
Max-Planck-Institut f\"ur Physik komplexer Systeme, N\"othnitzerstra{\ss}e 38,
D-01187, Dresden, Germany\\ 
}
\author{Andreas Buchleitner}
\affiliation{%
Max-Planck-Institut f\"ur Physik komplexer Systeme, N\"othnitzerstra{\ss}e 38,
D-01187, Dresden, Germany\\ 
}
\affiliation{
Albert-Ludwigs-Universit\"at Freiburg, Physikalisches Institut,
Hermann-Herder-Str. 3, D-79104 Freiburg, Germany\\
}
\author{Florian Mintert}
\affiliation{%
Max-Planck-Institut f\"ur Physik komplexer Systeme, N\"othnitzerstra{\ss}e 38,
D-01187, Dresden, Germany\\ 
}
\affiliation{
Albert-Ludwigs-Universit\"at Freiburg, Physikalisches Institut,
Hermann-Herder-Str. 3, D-79104 Freiburg, Germany\\
}
\affiliation{%
Department of Physics, Harvard University,
17 Oxford Street, Cambridge Massachusetts, USA}
\date{\today}%
\begin{abstract}
We present
an efficient experimental estimation of the multipartite entanglement of mixed
quantum states 
in terms of simple parity measurements.
\end{abstract}
\maketitle
\emph{Introduction.}--  Entanglement has been identified as a key resource for
quantum information processing tasks. Furthermore, it is clear that the
dramatic 
advantage in using quantum mechanical systems instead of 
classical ones to process
information emerges only in the limit of a very large 
number of system components. 
Taking into account all the 
extra qubits necessary for error correction, a quantum computer has to run on
quantum registers composed of at least several thousand qubits to outperform
its present-day classical counterpart.  
This explains the tremendous effort dedicated during 
the last few
years to the experimental production and coherent manipulation of
multiparticle entangled states  
of photons \cite{Eibl, Witnesses, Bowmeester,Pan, Pan2, Zhao}, ions
\cite{Roos, Sackett, Leibfried, Haeffner}, and in cavity QED devices
\cite{Raschenbeutel}. 

\par Also the experimental {\em quantification} of multipartite entanglement
has 
thus become a major issue of interest. 
In principle, such quantification can
be carried out through quantum state tomography \cite{White,Roos,Roos2},
\emph{i.e.}, the 
complete reconstruction of the state's density matrix via the
measurement of a complete set of observables, followed by the subsequent
evaluation of a valid entanglement measure. In practice, however, 
tomography rapidly saturates the available
resources and is thus no viable strategy 
under the perspective of scalability. 
Clear evidence of this is given by
the experimental characterization of genuine multiparticle entangled states of up to
eight ions \cite{Haeffner}:
Ten hours of data aquisition -- implementing 
measurements in $3^8=6561$  detection bases, each corresponding to a different
experimental setting --  
were followed by computationally expensive data processing, 
to reconstruct the eight-ion density matrix of the experimentally prepared
state. 
Therefore, full tomography of
entangled ion chains composed of more than only eight ions appears largely
impracticable. 

\par Quantum non-locality tests \cite{Eibl, Bowmeester, Pan, Pan2, Zhao,
 Raschenbeutel} and entanglement witnesses \cite{HoroWitness, Witnesses,
 Roos, Sackett, Leibfried, Haeffner, kiesel07, bovino05} provide alternative
 means to assess the  
degree of entanglement of a quantum state, and  
were used in several experiments.
Both these techniques require the measurement of only a
few observables, 
but allow to detect the entanglement of only a
small class of states. 
This implies that some a priori
knowledge on the state to be analyzed is necessary.
A simple entanglement measurement scheme for {\em arbitrary} mixed
states is therefore highly desirable.
 
\par
First steps in this direction were taken in \cite{Flo-Review,Boludo-Flo},
where multipartite concurrence \cite{Andre} was shown to be
directly accessible through projective measurements on two identically
prepared quantum states. This original approach
\cite{Flo-Review,Boludo-Flo} -- experimentally demonstrated for twin
photon entanglement \cite{Steve,Steve2} -- was restricted to the ideal case of
pure 
states, and a first generalization for mixed states was given in
\cite{Flo-Andreas}, yet 
applicable only for bipartite systems. Here we come up with the ultimate 
formulation of this approach to direct experimental entanglement estimation, 
for mixed states of quantum systems with an {\em arbitrary} number of
constituents. Our procedure, based on local parity
measurements, features excellent scaling properties: the number of required 
observables -- which can all be probed in one single experimental setting --
is equal to the number of subsystems. 

\par \emph{The observable lower bound.}--
Consider two copies of an arbitrary
mixed state $\varrho$ of an $N$-partite quantum system with Hilbert space
${\cal H}$. 
We introduce an observable $V$ such that 
\begin{equation}
\label{lowerbound}
C^{2}(\varrho)\geq\mbox{Tr}\big[\varrho\otimes\varrho\ V \big]\ , 
\end{equation}
i.e., that allows us to experimentally bound the concurrence of the state
from below. The Hermitian operator $V$ acts on the composite Hilbert space
associated with the two-fold copy of the system ${\cal H}\otimes {\cal H}$, and
has the two following remarkable properties:  \emph{ (i)} it can be detected
through  \emph{projective measurements of only $N$ two-particle observables},
and  \emph{ (ii)} \emph{a single experimental setting is required} throughout
the detection process.

\par
\begin{figure}[t]
\begin{center}
\includegraphics[width=1\linewidth]{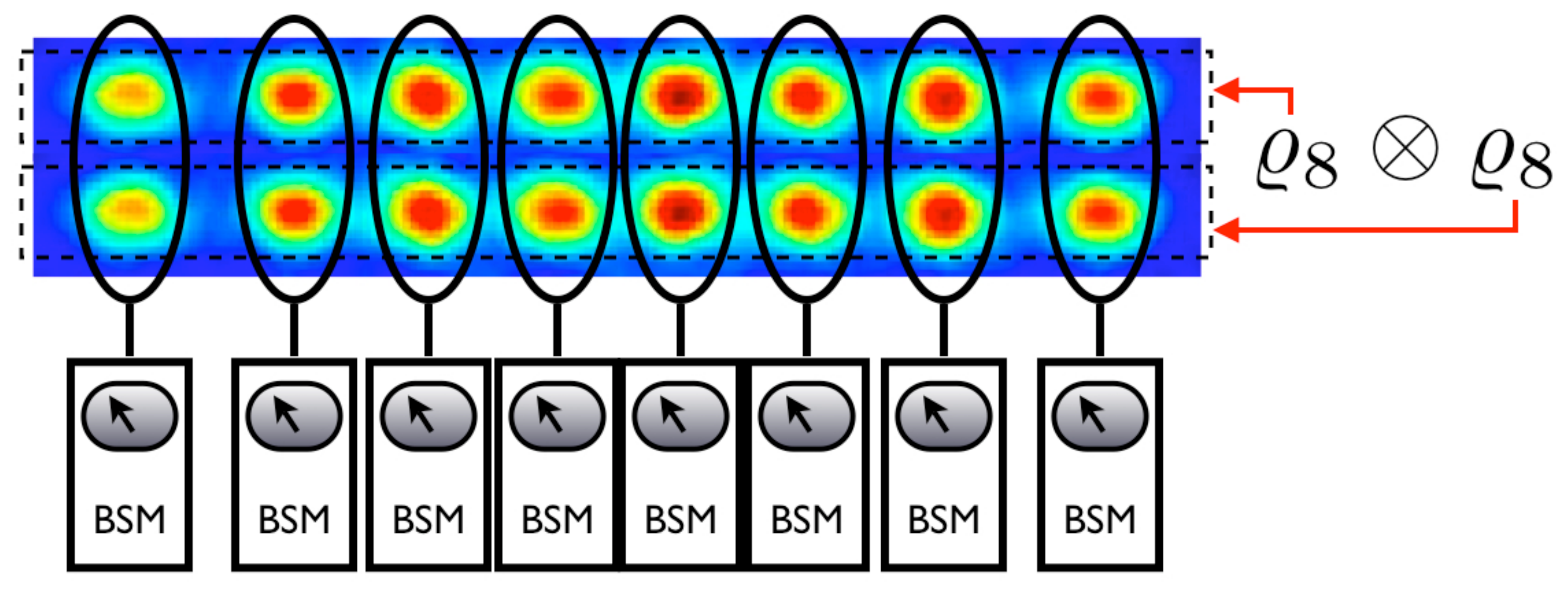}
\caption{
\label{8ion}
An eight-qubit state $\varrho$, together with its copy,
is encoded, for example, in strings of two-level ions and is subject to
Bell-state measurements (BSMs) on each pair. The entanglement of $\varrho$
is obtained from the joint probabilities of appearance of singlets and
triplets.} 
\end{center}
\end{figure}

 The required two-particle measurements are 
simultaneous parity
measurements on each particle and its copy.
In each run of the experiment,
measuring the local parity state of all $N$ pairs defines an event in
which each pair is projected onto either a symmetric or an antisymmetric
state. From all possible events we distinguish three types:
\begin{itemize}
\item[{\it (i)}] the entire system
together with its copy is projected onto a globally symmetric state --
symmetric 
with respect only to the exchange of both copies of the entire system;
\item[{\it (ii)}] the entire system and copy are projected onto a globally
  antisymmetric 
state -- antisymmetric with respect only to the exchange of both copies of the
entire system; or,
\item[{\it (iii)}] system and copy are projected onto a full
locally symmetric state in which all $N$ particle-copy pairs are
simultaneously found to be in a symmetric state
(which is a particular case of \emph{(i)}).
\end{itemize}
The probabilities of these three events suffice to obtain the expectation
value of $V$, 
as described below.
The measurement protocol is sketched in Fig. \ref{8ion}, for 
two strings 
of eight ions, reminiscent of the experimental situation in \cite{Haeffner}.
Each particle, together with its counterpart in the copy, is subject  to a
local parity measurement, 
which reduces to a Bell-state measurement, since the particles are qubits in
this example. 
Recording the abundance of singlets in the string of eight ion pairs allows 
to infer the probabilities of the three above events immediately. 

\par From a more technical point of view, the system's Hilbert
space ${\cal H}$ is a tensor product 
${\cal H}\equiv {\cal H}_{1}\otimes {\cal H}_{2}\otimes ... \otimes {\cal
  H}_{N}$ 
of the single particle Hilbert spaces 
${\cal H}_{i}$, $1\leq i \leq N$.
The symmetric and antisymmetric subspaces
${\cal H}_{i}\odot{\cal H}_{i}$ and ${\cal H}_{i}\wedge{\cal H}_{i}$
of the Hilbert space ${\cal H}_{i}\otimes{\cal H}_{i}$ of 
two copies of the $i$-th single-particle subsystem are defined as the
subspaces spanned by  
all states that acquire a phase shift of $0$ or $\pi$, respectively,  
upon exchange of the
single-particle copies. These two subspaces are associated with the
local two-particle projectors $P^{i}_{+}$ and $P^{i}_{-}=
1-P^{i}_{+}$. 
The globally symmetric and antisymmetric subspaces
${\cal H}\odot{\cal H}$ and ${\cal H}\wedge{\cal H}$ 
are the subspaces of all states that are symmetric and antisymmetric with
respect to 
the exchange of two copies of the {\em entire} system, and not only of some
subsystems, and are in turn associated with 
the global projectors ${\bf P}_{+}$ and  ${\bf P}_{-}= {\bf 1}-{\bf P}_{+}$.
In terms of these, our 
observable can be explicitly expressed as 
\begin{eqnarray}
\label{desiredobservable}
V=4\Big({\bf P}_{+}-P^{1}_{+}\otimes\hdots\otimes P^{N}_{+}-(1-2^{1-N}){\bf P}_{-}\Big) . 
\label{newobs}
\end{eqnarray}
Since the symmetric (antisymmetric) global projector
${\bf P}_{+}$ (${\bf P}_{-}$) can be decomposed into a sum 
of all products of $N$ local projectors with an even (odd)
number of antisymmetric local projectors, 
it suffices to measure the parity of
the $N$ pairs of copies to reconstruct
$\mbox{Tr}\big[\varrho\otimes\varrho\ V\big]$. 

\par Finally, it is important to note that 
$V$, as defined in~(\ref{desiredobservable}), has an equivalent 
interpretation to that of its bipartite analogue \cite{Flo-Andreas}, just
with a much more intricate combinatorial structure: 
the expectation value of
$A=4\left({\bf P}_{+}-P^{1}_{+}\otimes\hdots\otimes P^{N}_{+}\right)$
yields the concurrence of pure states \cite{Boludo-Flo}.
For a general state $\varrho$, however, a positive
expectation value of $A$ can have two causes: entanglement or mixedness of
$\varrho$. 
In turn, the operator ${\bf P}_{-}$ quantifies the degree of mixing of
$\varrho$: $1-\mbox{Tr}[\varrho^2]=2\mbox{Tr}[\varrho\otimes\varrho\ {\bf
  P}_{-}]$.  
The linear combination of $A$ and ${\bf P}_-$ in (\ref{newobs}) therefore
rescales the expectation value of $A$ with respect to the state's intrinsic
impurity, and thus provides an estimate of the inscribed entanglement 
through a lower bound of multipartite concurrence, as elaborated in the appendix. 

\begin{figure}[b]
\begin{center}
\includegraphics[width=1\linewidth]{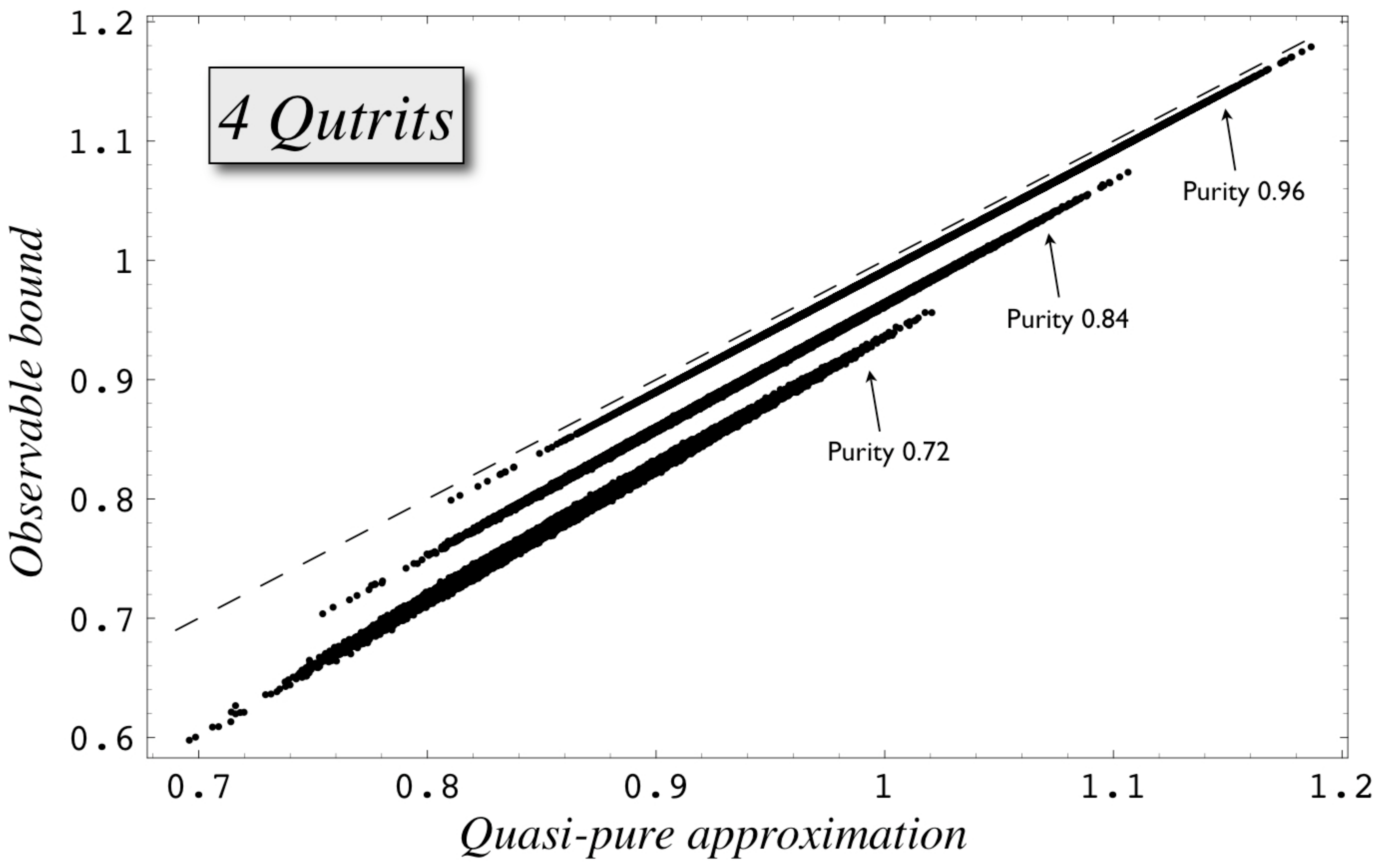}
\caption{
\label{4Qutrit}
Observable lower bound versus concurrence, in quasi-pure approximation for
$10^5$ four-qutrit density matrices with strong, intermediate, and weak
mixing. The dashed line indicates equality of our present measurable bound and
of entanglement in quasi-pure approximation. The tightness of
the observable bound is excellent for strongly entangled or weakly mixed states, 
but remains surprisingly good even for strong mixing.} 
\end{center}
\end{figure}
\par
\begin{figure}[t]
\begin{center}
\includegraphics[width=1\linewidth]{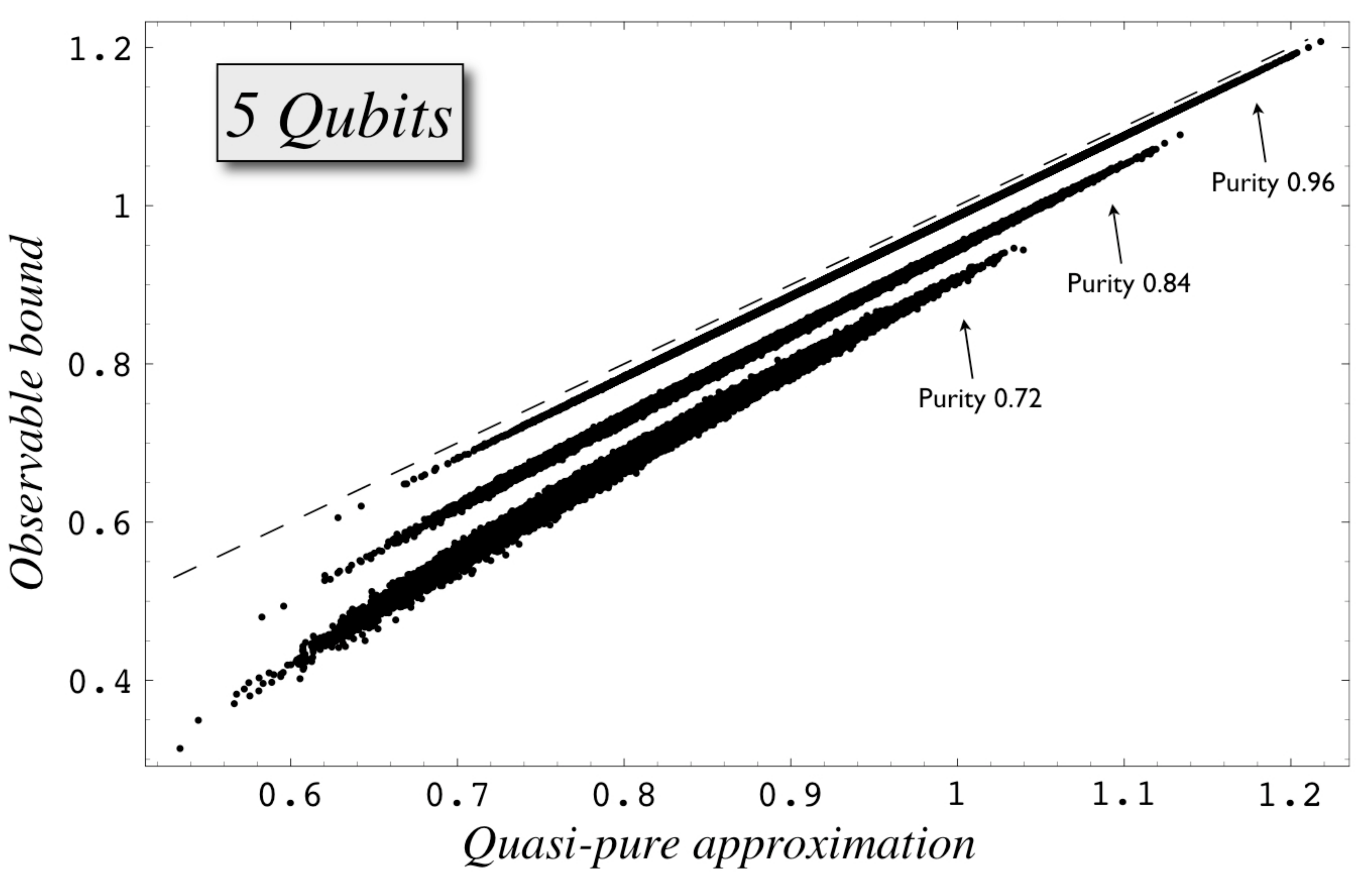}
\caption{
\label{5Qubit}
Same as Fig.~2, now for $10^5$ five-qubit density matrices at various levels
of purity. 
Once more, the tightness of
the observable bound is excellent for strongly entangled or weakly mixed
states, 
and still surprisingly good for strongly mixed states.} 
\end{center}
\end{figure}

\emph{Tightness of the bound}.-- 
Let us finally test the tightness of the observable bound on 
mixed random states. In Figs.~\ref{4Qutrit} and \ref{5Qubit} we 
plot the expectation value of the operator~(\ref{desiredobservable}), versus 
concurrence in quasi-pure approximation \cite{FloPRA}
(which is known to yield very good approximations for weakly mixed states),
for $10^5$ random states of 4-qutrit and 5-qubit systems, respectively, and
for different degrees of mixing. 
Mixed states of different purity were obtained by acting with 
the generalized depolarizing channel (which essentially mixes a pure state
with the identity) \cite{Boludo-Nacho} onto 
$10^5$ random pure states, 
for three different coupling strengths.

\par As spelled out by the comparison in Figs.~\ref{4Qutrit} and
\ref{5Qubit}, the observable bound is hardly weaker than the quasi-pure
approximation. 
In fact, the comparison is excellent for weakly mixed or highly entangled
states. 
On the other hand, for some  very strongly mixed or very weakly entangled
states other techniques involving few measurements, such as `tailored
witnesses' \cite{Guehne}, may be used to improve the  tightness of the
entanglement estimation if  some  a priori knowledge of the state is
available. 
The expectation value of  (\ref{desiredobservable}), however, provides a
\emph{directly observable} 
non-trivial bound for \emph{any unknown} multipartite state's concurrence. 
\par\emph{Conclusions}.--
We have derived a general lower bound for the entanglement of 
mixed quantum states, which provides  
a hierarchy of observable entanglement measures. As such, our result has the
essential virtue of scalability for unknown, multipartite mixed quantum states in
arbitrary finite dimensions. 
Given a two-fold copy of the state to be analyzed, our bounds are
experimentally accessible, 
with linear scaling of the experimental overhead with the number of system constituents. 
While derived for a specific type of multipartite concurrence
\cite{Andre,rafal07}, 
equivalent expressions can be found for other
observable multipartite concurrences with the same algebraic
structure \cite{Flo-Review,rafal07}.
This defines a versatile
toolbox for the experimental probing of quantum correlations inscribed into
ever larger 
multicomponent quantum systems, an essential prerequisite for
scaling up quantum information technology.
\par\emph{Appendix}.--
Here we prove that the 
observable defined in (\ref{desiredobservable}) satisfies
Eq.~(\ref{lowerbound}), for any state $\varrho$:
The concurrence of $\varrho$ is given by the convex roof \cite{Bennett}
$C(\varrho)= \inf\sum_j C(\psi_{j})$,
{\it i.e.} the minimal average concurrence over all
(subnormalized) pure-state decompositions
$\varrho=\sum_{j}\ket{\psi_{j}}\bra{\psi_{j}}|$.
If  $\sum_{jk} C(\psi_{j})C(\psi_{k})\geq
\mbox{Tr}\big[\varrho\otimes\varrho\ V \big]=
\sum_{jk}\bra{\psi_{j}}|\otimes\bra{\psi_{k}}|V\ket{\psi_{j}}\otimes\ket{\psi_{k}}$  
holds for \emph{all} decompositions  $\{\ket{\psi_{j}}\}$, then it also
holds for the optimal convex-roof decomposition, and inequality
(\ref{lowerbound}) is automatically satisfied.
Therefore, we seek $V$ such that
\begin{equation}
\label{lowerboundtemr2term}
C(\psi)C(\phi)\geq\bra{\psi}|\otimes\bra{\phi}|V\ket{\psi}\otimes\ket{\phi}\ 
\end{equation}
holds for any two arbitrary pure states
$\ket{\psi},\ket{\phi}\in{\cal H}$.
Such an observable is known for the bipartite concurrence $c$:
$v=4\big({\bf P}_{+}-P^{1}_{+}\otimes P^{2}_{+}-\frac{1}{2}(P^{1}_{-}\otimes
P^{2}_{+}+P^{1}_{+}\otimes P^{2}_{-})\big)$ \cite{Flo-Andreas}.
Now, we can make use of the fact that the $N$-partite concurrence can be
decomposed into bipartite terms as 
\begin{equation}
C(\Psi)=2^{1-N/2}\sqrt{\sum_ic^2_i(\Psi)}\ ,
\end{equation}
where the sum is taken over the bipartite concurrencies $c_i$ corresponding to
each subdivision of the entire 
system into two subsystems.
This allows us to bound our quantity of interest from below as
\begin{eqnarray}
C(\Psi)C(\Phi)&=&2^{2-N}\sqrt{\sum_ic^2_i(\Psi)}\sqrt{\sum_ic^2_i(\Phi)}\\
&\ge&
2^{2-N}\sum_ic_i(\Psi)c_i(\Phi)\\
\label{last}
&\ge&
2^{2-N}\sum_i\bra{\psi}|\otimes\bra{\phi}|v_i\ket{\psi}\otimes\ket{\phi}\ ,
\end{eqnarray}
where we made use of the Cauchy-Schwarz inequality $\sqrt{\sum_i
  x_{i}^{2}}\sqrt{\sum_i y_{i}^{2}}\geq\sum_i x_{i} y_{i}$ and the above
knowledge on bipartite systems. 
It is now a matter of straightforward algebraic gymnastics to
show that $V=\sum_iv_i$, what finishes the proof of Eq.~(\ref{lowerbound}).

\par \emph{Acknowledgements}.--
We gratefully acknowledge substantial financial support within the
PROBRAL program of the German Academic Exchange Serivce (DAAD), through
the Feodor Lynen program of the Alexander von Humboldt Foundation (AvH),
as well as by FAPERJ, CAPES, and the Brazilian Millenium Institute for
Quantum Information. 


\end{document}